# Inelastic tunneling in a double quantum dot coupled to a bosonic environment

Toshimasa Fujisawa[1,*], Wilfred G. van der Wiel[1,2], Leo P. Kouwenhoven[2]

[1]*NTT Basic Research Laboratories, 3-1 Morinosato-Wakamiya, Atsugi, 243-0198, Japan.*
[2]*Department of Applied Physics and DIMES, Delft University of Technology, 2600 GA Delft, Netherlands.*

**Abstract**
Coupling a quantum system to a bosonic environment always give rise to inelastic processes, which reduce the coherency of the system. We measure energy dependent rates for inelastic tunneling processes in a fully controllable two-level system of a double quantum dot. The emission and absorption rates are well reproduced by Einstein's coefficients, which relate to the spontaneous emission rate. The inelastic tunneling rate can be comparable to the elastic tunneling rate if the boson occupation number becomes large. In the specific semiconductor double dot, the energy dependence of the inelastic rate suggests that acoustic phonons are coupled to the double dot piezoelectrically.

**Introduction**

Semiconductor quantum dots, which are often referred to as artificial atoms, contain well-defined discrete energy states [1,2]. If two of these quantum dots are coupled, an artificial two-level system shows controllable wave-particle duality [3,4]. When one extra electron is added into the double quantum dot, a classical particle picture represents that the electron occupies either one quantum dot or the other. A quantum wave picture represents that the electron may spread over the two dots as a superposition of the charge states. Multiple gated structures for the double dot allow us to independently control their energy levels, the coupling strength between the dots, as well as the tunneling rate to the leads. A dc bias voltage and an ac (microwave) voltage can be used to modulate the electron occupation of the states [5]. This controllability allows us to study the manipulation of quantum states for future quantum logic gates [6]. Nonadiabatic control of the states could be used to create a coherent charge oscillation for a desired time [7].

In general, a controllable quantum system may have decoherency problems due to the coupling to its environment [8]. Inelastic processes accompanied by energy dissipation are crucial for the dynamics of the quantum system. A double quantum dot has a tunable energy selectivity to its environment, since the system emits and absorbs an energy equal to the energy spacing of the two levels [4]. One can measure the energy dependent coupling to the environment without any external spectrometer.

In this paper we study inelastic tunneling in a double dot system for different tunneling regimes and temperatures [9]. We provide a full data set of the double quantum dot coupled to the lattice environment. We can relate the emission and absorption rates to the Einstein coefficients over the full energy and temperature range we studied. At the lowest temperature (23 mK), we directly measure the energy dependent spontaneous emission rate. In our specific semiconductor device, the lattice environment gives rise to piezoelectric electron - (acoustic) phonon interaction.

**Tunable two-level system**

The double quantum dot investigated here, is fabricated in an AlGaAs/GaAs two-dimensional electron gas (2DEG) using focused ion beam im-plantation and fine Schottky gates (Fig. 1a) [4,5,9]. Following measurements were performed in a dilution refrigerator at a lattice temperature, $T$, between the base temperature of 23 mK and 300 mK. The electron temperature of the device at $T = 23$ mK is about 50 mK. The magnetic field of 5 or 7.5 T is applied to reduce co-tunneling processes.

Application of negative voltages, $V_{GL}$, $V_{GC}$, and $V_{GR}$, on the respective gates, $G_L$, $G_C$, and $G_R$, forms tunneling barriers in the 2DEG. The central gate $G_C$ is used to change the tunneling coupling, $T_c$, between the two dots. The outer gates, $G_L$ and $G_R$, lift the electrostatic potential of the two dots, L and R, and change the number of electrons in the dots ($n$, $m$).

We consider the ground state, $E_L$, of the $n$-electron system in dot L and the ground state, $E_R$, of the $m$-electron system in dot R. Since the electron occupations of the two dots are correlated due to the electrostatic coupling, the two states can not be occupied simultaneously [5]. This is effectively a *two-level system*. Other states such as excited and bound states can be ignored as long as the bias voltage and the thermal energy are smaller than the required excitation energy.

The gray scale plot of the current in the $V_{GR}$-$V_{GL}$ plain (Fig. 1(b)) was taken near an electron-like resonance [5], in which an electron sequentially tunnels three barriers in the normal order to the net electron flow. The fine tuning voltage on the gates $G_L$ and $G_R$ effectively lifts the energy states, $E_L$ and $E_R$, in the dots L and R, respectively, which is illustrated by arrows in the figure. Finite current is observed when both states are located in the transport window, which ensures no co-tunneling current (<10 fA). The dominant current, which is seen as a sharp line along the upper right direction in Fig. 1(b), is the elastic tunneling between $E_L$ and $E_R$, hence $\varepsilon \equiv E_L - E_R = 0$ (Fig. 1(d)). In addition to the elastic tunneling, the inelastic tunneling from the higher level $E_L$ to the lower level $E_R$ appears in the upper-left triangle region, $\mu_S > E_L > E_R > \mu_D$ (Fig. 1(c)). Here $\mu_S$ and $\mu_D$ are the Fermi energies of the source and the drain, respectively. Absorption may occur in the lower right triangle, $\mu_S > E_R > E_L > \mu_D$ (Fig. 1e), but no current contribution from the absorption is observed at the lowest temperature.

**Elastic and inelastic current spectrum**
In the following experiments, we swept $V_{GL}$ and $V_{GR}$ simultaneously in opposite directions to change the energy difference of the two levels, $\varepsilon \equiv E_L - E_R$ (see the arrow $\varepsilon$ in Fig. 1b). We can measure the absorption, elastic and emission spectrum in a single trace. The inelastic current can be measured until one of the states is out of the transport window. Fig. 2(a) shows the current spectrum for positive and negative bias voltage. The inelastic current appears in positive $\varepsilon$ at $V_{sd} > 0$ and in negative $\varepsilon$ at $V_{sd} < 0$, where emission occurs. Some step-like structures in the inelastic part are reproduced for both bias directions, which will be discussed later.

For the quantitative analysis, we decomposed the current into a symmetric part $I_{el}(\varepsilon) = I_{el}(-\varepsilon)$ (dashed curve) and the remaining asymmetric part $I_{inel}(\varepsilon>0)$ (dotted-dashed curve) as shown in the inset of Fig. 3a. The elastic current should have a Lorentzian line shape [3,10]

$$I_{el}(\varepsilon) = eT_c^2 \Gamma_R / \{T_c^2(2 + \Gamma_R/\Gamma_L) + \Gamma_R^2/4 + (\varepsilon/\hbar)^2\}, \quad (1)$$

where $\Gamma_L$ and $\Gamma_R$ are the tunneling rates for the left (incoming) and right (outgoing) barrier for $V_{sd} > 0$, and they should be swapped in the expression for negative $V_{sd}$. Note that $\Gamma_R$ and $\Gamma_L$ appear in an asymmetric way, because the incoming tunneling builds up the single electron states and the outgoing tunneling breaks the states. Thus, measurements of the Lorentzian line shape for both positive and negative biases determine all the tunneling parameters, $T_c$, $\Gamma_L$ and $\Gamma_R$. Fig. 2(c) shows a logarithmic-linear plot of the current spectrum, whose elastic part is well reproduced by a Lorentzian curve.

The inelastic current is nonzero over the range of 100 μeV, despite the thermal energy $kT$(23 mK) = 2 μeV being much smaller. The inelastic process should be sequential tunneling of the three barriers, which yields

$$I_{inel}(\varepsilon) = e/(\Gamma_L^{-1} + \Gamma_{inel}^{-1}(\varepsilon) + \Gamma_R^{-1}). \quad (2)$$

When $\Gamma_R$ and $\Gamma_L$ are larger than the inelastic tunneling rate $\Gamma_{inel}$, the current becomes $I_{inel}(\varepsilon) = e\Gamma_{inel}(\varepsilon)$.

The temperature dependence of the inelastic current is shown in Fig. 3(a). A higher temperature $T$ enhances the inelastic current on both the emission and the absorption side. In the following analysis, we assume boson statistics for the degrees of freedom in the environment. The average occupation number $<n>$ of environmental modes is given by the Bose-Einstein distribution function: $<n> =$

$1/(e^{\varepsilon/kT} - 1)$.

The rates for absorption, $W_a$, and emission, $W_e$, can be expressed very generally by $W_a = B_a\rho$ and $W_e = A + B_e\rho$, where the Einstein coefficients stand for spontaneous emission ($A$), stimulated emission ($B_e$) and absorption ($B_a$), and $\rho$ is the energy density [11]. From the Einstein relations, $B_a = B_e = A<n>/\rho$, we obtain:

$$\Gamma_i(\varepsilon<0) = W_a(\varepsilon) = <n> A(-\varepsilon) \qquad (3)$$
$$\Gamma_i(\varepsilon>0) = W_e(\varepsilon) = (<n>+1) A(-\varepsilon).$$

We have tested this relation in our quantum dot system. The normalized rate $W_e(\varepsilon)/A(\varepsilon)$ and $W_a(\varepsilon)/A(-\varepsilon)$ are plotted versus $kT/|\varepsilon|$ for various $\varepsilon$ (18, 24, 40, 60, and 80 μeV) and T (23, 75, 100, 125, 150 and 200 mK). The measured data follow the $<n>$ and $<n>+1$ curves without any fitting parameters as shown in Fig. 3(b). For temperatures above 200 mK, the emission and absorption rates increase faster than predicted, which may be due to thermally excited electrons. Nevertheless, the inelastic tunneling is well described by spontaneous emission that is caused by vacuum fluctuations in the environment.

Another example of the temperature dependence is plotted in Fig. 3(c), in which we choose the tunneling barriers to give the minimum peak width of 4 μeV at 23 mK. Increasing temperature, however, gives rise to stimulated processes and significantly broadens the current peak. The full width at half the maximum (FWHM) of the peak is plotted versus temperature in Fig. 3(d). The FWHM is obviously smaller than 3.5 $kT$ (dashed line) which is the peak width in the case of single quantum dots [12], and can be even smaller than $kT$ (solid line). Nevertheless, the FWHM has a finite temperature dependence and is not completely determined by the tunneling coupling. The stimulated emission and absorption rates are proportional to $<n>$, and can be comparable to the elastic rate if $|\varepsilon| < kT$.

In our measurement, the two levels, $E_L$ and $E_R$, are more spatially separated if $|\varepsilon|$ is increased. The overlap depends on $(T_c/\varepsilon)^2$, thus the spontaneous emission rate for the double quantum dot should follow

$$A(\varepsilon) \sim (T_c/\varepsilon)^2 J(\varepsilon), \qquad (4)$$

where the spectral function $J(\varepsilon)$ describes the energy dependent coupling to the environment. We have measured the $T_c$ dependence in Fig. 2(b). We apply more positive voltage to the central gate to increase $T_c$, and more negative voltages to the outer gates to compensate for the energy shifts. In the weak coupling regime where $\Gamma_R, \Gamma_L > T_c$, the elastic part of the current is well reproduced by a Lorentzian line shape (see Fig 2(c)). We obtain rough estimates of $\Gamma_R$, $\Gamma_L$ and $T_c$ from the fitting with Eq. 1, and $\Gamma_{inel}(\varepsilon)$ from the inelastic current using Eq. 2. We found that with these fitting values, the inelastic current scales as $T_c^\alpha$ with an exponential factor $\alpha$ of 2.5 ~ 3, which is somewhat larger than expected. For the strong coupling case ($\Gamma_R, \Gamma_L < T_c$), the current is rather restricted by the outer barriers. We found a significant deviation from the Lorentzian line shape for small $\varepsilon$, implying that the elastic and inelastic rates can become of the same order. This effect may form a significant limitation for the coherence time in coupled quantum dot devices. For the largest coupling, we obtain a saturated current where the elastic current peak can no longer be distinguished.

**Coupling to acoustic phonons**

The possible bosonic environment coupled to the double quantum dot might be formed by phonons, plasmons, or photons. Acoustic phonons are most likely the bosons coupled to the double dot [9]. In solid-state systems the electron - acoustic phonon interaction is strong. The typical emission rate in a single quantum dot at zero temperature can be $10^8$ - $10^{10}$ s$^{-1}$, due to deformation or piezoelectric interaction [13]. Since the wave functions in a double dot system are spatially separated, the spontaneous emission rate is lower than that, but still high enough to contribute a current of a few pA.

For the energy regime of our interest, 10 - 100 μeV, acoustic phonons have linear dispersion with a constant sound velocity. We consider 3D bulk GaAs phonons and 2D surface acoustic waves, since the double dot is located in GaAs, but very close to the surface. The dimensionality of the phonons gives different densities of states, $\rho^{3D} \sim \varepsilon^2$ and $\rho^{2D} \sim \varepsilon$. If the double dot is coupled to phonons with a deformation type coupling, the dilatation of the lattice, $\Delta$, shifts electrostatic potentials of the dots, $\delta\varepsilon \sim E_{def}\Delta$. Here $E_{def}$ is the deformation potential. In the case of piezoelectric coupling existing in polar materials, the dilatation gives the electric field and thus, $\delta\varepsilon \sim e_{pz}\Delta/k$. Here $e_{pz}$ is the piezoelectric constant and $k$ is the wave vector of the phonons [14]. Since the typical wavelength of the phonons is comparable to the size of the double dot, the spectral function, $J(\varepsilon)$, will have a complicated energy dependence. Neglecting the oscillating term, we obtain an overall energy dependence, $J(\varepsilon) \sim \varepsilon^D$ for deformation interaction and $\sim \varepsilon^{D-2}$ for piezoelectric interaction. The dimensionality, $D$, is 2 or 3 in our case. A logarithmic plot of the inelastic current (Fig. 4) indicates an energy dependent rate of $\varepsilon^{-1}$ or $\varepsilon^{-2}$. Assuming the coupling dependence of Eq. 4, $J(\varepsilon)$ is either linearly dependent of $\varepsilon$ or constant, implying piezoelectric interaction of 2D or 3D phonons.

The step-like structures in the inelastic rate might be due to the oscillating term in the interaction. The maximum strength should appear when half the phonon wavelength, $\lambda/2$, equals the spacing of the two dots, $L$, i.e. at the resonance energy

$$\varepsilon_r = hs/2L \approx hs/\ell. \quad (5)$$

where the geometric size, $\ell$, between the outer gates defines the location of the dots, $L \approx \ell/2$. The sound velocity, $s$, is $s^{2D} = 2800$ m/s for surface acoustic waves or $s^{3D} = 4800$ m/s for GaAs bulk phonons. On the other hand, surface gates might modify the surface vibration mode, which affects on the electron transport if the inelastic process dominated by 2D phonons. The characteristic resonance should appear at $\varepsilon_r \approx hs/\ell$ in our geometry, which is the same as Eq. 5. The inset shows the energy of the step-like structures in the spectrum (circles) and the condition of Eq. 5 (lines), which agree well. Although only two samples have been investigated, we believe that a kind of phonon resonance appears in the electronic transport. We expect that the interaction with phonons can be controlled in phonon cavities, as is often realized in photon cavities.

**Summary**

We have investigated inelastic tunneling between spatially separated discrete levels in double quantum dots. The lattice environment gives a significant contribution to inelastic processes in a polar solid state system.

**Acknowledgement**

We thank R. Aguado, L. Glazman, J. Mooij, T.H. Oosterkamp, and S. Tarucha for help and discussions. L.P.K. acknowledges financial support from the Specially Promoted Research, Grant-in-Aid for Scientific Research, from the Ministry of Education, Science and Culture in Japan, from the Dutch Organization for Fundamental Research on Mater (FOM), from the NEDO joint research program (NTDP-98), and from the EU via a TMR network.

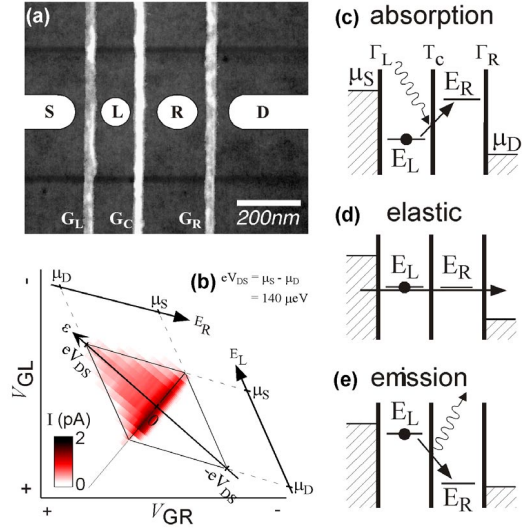

Fig. 1. (a) SEM picture of the double quantum dot defined in the 2DEG of a GaAs/AlGaAs hetero structure. The dark horizontal lines are the focused ion beam pattern, which appears as spattered surface with more than 100 times higher ion dose of that used in the real sample. The bright vertical lines are Schottky gates. The two quantum dots, $L$ and $R$, respectively, contain ~15 and ~25 electrons; charging energies are ~4 and ~1 meV; and the measured average spacing between single-particle states are ~0.5 and ~0.25 meV. (b) Gray scale plot of the current through the double dot in the $V_{GL}$-$V_{GR}$ plane. The ground state of the left and the right dot, $E_L$ and $E_R$, are varied as indicated by the arrows. The arrow ε indicates the voltage-vector used in the current spectroscopy. (c, d, and e) Energy diagrams of the double dot for the tunnel situations: Absorption, elastic tunneling and emission. The continuum of electron states in the leads is filled up to the Fermi energies $\mu_S$ and $\mu_D$. The voltage $V_{sd}$ between the leads opens a transport window of size: $eV_{sd} = \mu_S - \mu_D$.

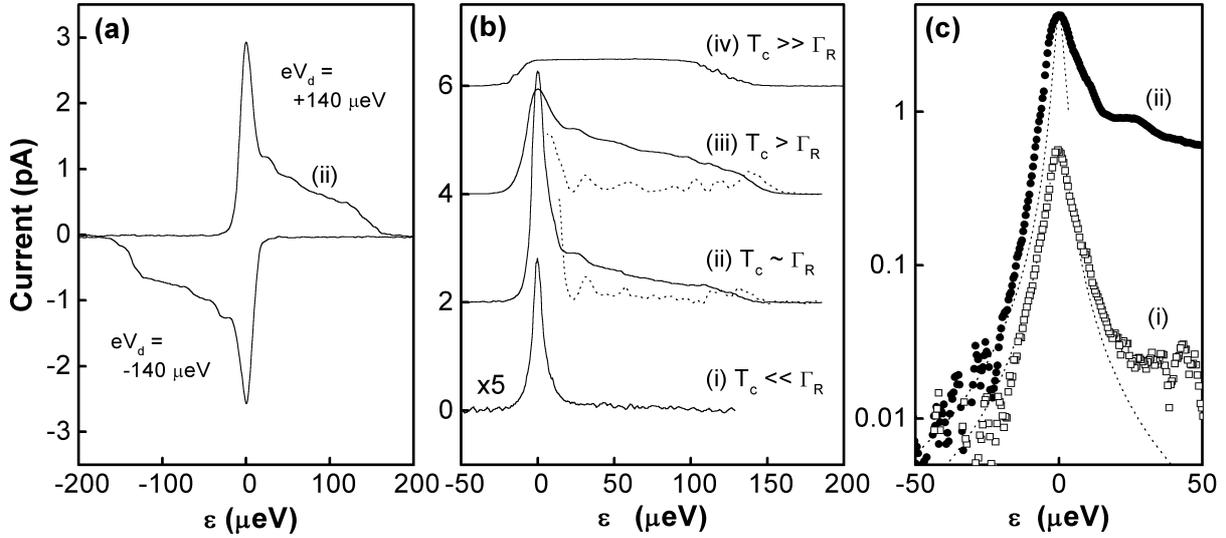

Fig. 2 (a) Typical current profiles for positive and negative $V_{sd}$, if the gate voltages are simultaneously swept in opposite directions such that we change the energy difference ε (see the arrow ε in Fig. 1(b)). (b) Current spectrum for different coupling energies at 23 mK. The curves have an offset, and curve (i) is multiplied by 5. Rough estimates for the coupling energies are: (i) $hT_c$ (~ 0.1 μeV) << $h\Gamma_R$ (~ 10 μeV), (ii) $hT_c$ ~ $h\Gamma_R$ (~ 1 μeV), (iii) $hT_c$ > $h\Gamma_R$ (~ 0.1 μeV), and (iv) $hT_c$ >> $h\Gamma_R$ (~ 0.01 μeV), and $\Gamma_L$ > $\Gamma_R$ for all curves. Dashed curves for (ii) and (iii) are $-dI(\varepsilon)/d\varepsilon$ curves in arbitrary units. (c) Logarithmic-linear plots for (i) and (ii). The dashed lines are Lorentzian fits.

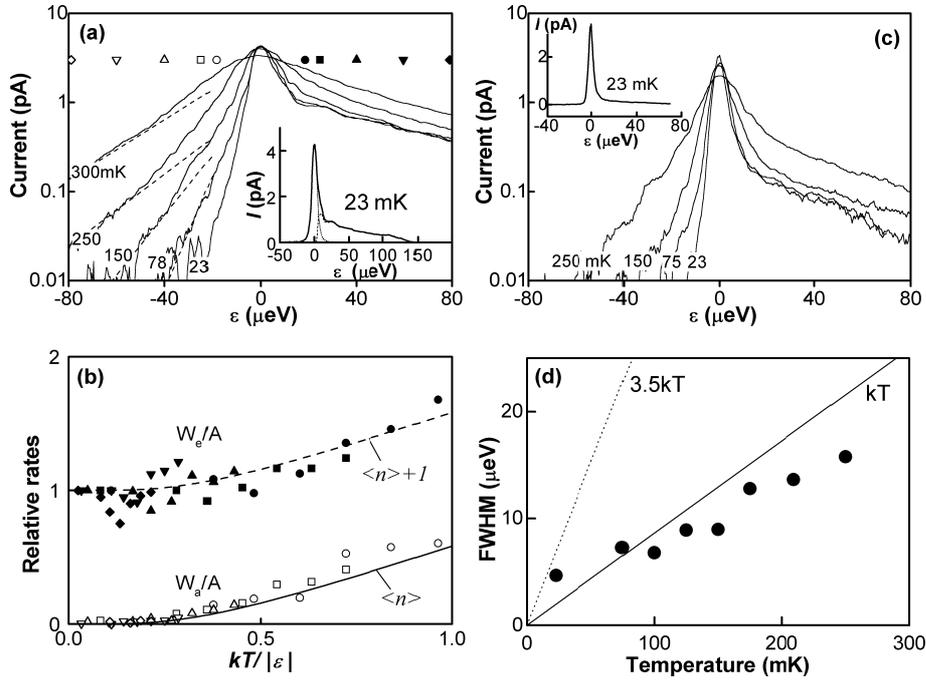

Fig. 3 (a) Temperature dependence of the current versus $\varepsilon$ for $hT_c \sim h\Gamma_R \sim 1$ μeV and $h\Gamma_L \sim 5$ μeV. The dashed lines indicate an exponential dependence, $e^{\varepsilon/kT}$. The inset shows the linear plot at 23 mK and the decomposition into the symmetric elastic part (dotted) and the inelastic part (dashed). (b) The absorption rate $W_a$ (open symbols) and emission rate $W_e$ (closed symbols) normalized by the spontaneous emission rate A versus $|kT/\varepsilon|$. Circles, squares, upper- and lower-triangles, and diamonds are taken at $|\varepsilon|$ = 18, 24, 40, 60, and 80 μeV, respectively (see also symbols in (a)). The solid line indicates the Bose-Einstein distribution, $<n>$, whereas the dashed line shows $<n>+1$. (c) Temperature dependence of the current versus ε for $hT_c \sim 0.5$ μeV and $h\Gamma_R \sim 3$ μeV. The inset is the linear plot at 23 mK. (d) Temperature dependence of the full width at half the maximum (FWHM) of the current peak in (c). Solid and dashed lines indicate $kT$ and $3.5kT$.

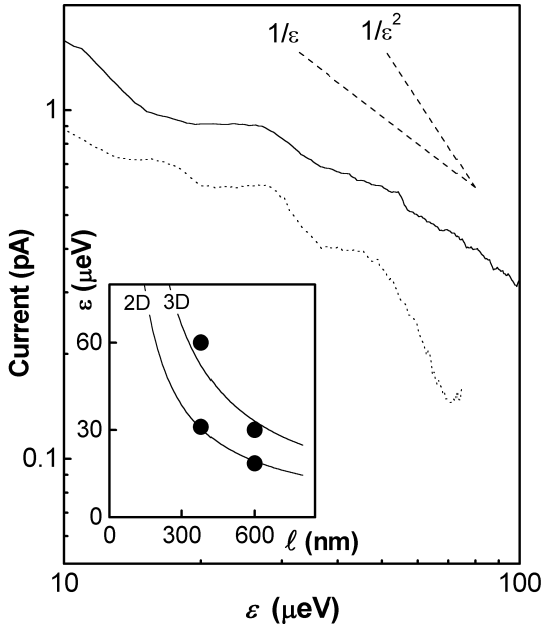

Fig. 4. Log-log plots of the emission spectrum for two different samples. The sold line is taken on the FIB sample ($\ell$ = 380 nm) in Fig. 1(a). The dotted line is taken on a surface gate sample with a distance between left and right barriers of $\ell$ = 600 nm [3]. The dashed lines indicate a $1/\varepsilon$ and $1/\varepsilon^2$ dependence expected for piezoelectric interaction with 3D and 2D phonons, respectively. The inset shows the energy position of the step-like structure, defined at the maximum in $-dI/d\varepsilon$ curves (see also Fig. 2(b)). The solid lines are expected phonon resonance condition for 3D ($s^{3D}$ = 4800 m/s) and 2D ($s^{2D}$ = 2800 m/s) sound velocity.